\documentclass[aps,prd,preprint,tightenlines,nofootinbib]{revtex4}
\usepackage{bm}
\newcommand{\etal}{{\it et al.}}
\newcommand{\Rprod}{R_{\mathrm{prod}}}
\begin{document}
\title{Ratio of Branching Fractions for $\chi_{cJ} \to \gamma J/\psi$}

\author{R.~S.~Galik}
\author{B.~K.~Heltsley}
\author{H.~Mahlke}
\affiliation{Cornell University, Ithaca, New York 14853}

\date{August 24, 2006}

\begin{abstract} 
One important quarkonia result from the Tevatron experiments 
is the ratio of direct inclusive
production of $\chi_{c1}$ to $\chi_{c2}$,
which is most readily measured using $\chi_{cJ}\to\gamma J/\psi$ decays.  This
note uses CLEO publications to 
obtain a ratio of these radiative
branching fractions,
${\cal B}(\chi_{c1}\to\gamma J/\psi)/{\cal B}(\chi_{c2}\to\gamma J/\psi) =
1.91 \pm 0.10$,
using cancelation in systematic uncertainties not available
in the Particle Data Group listings.
\end{abstract}

\maketitle

One important quarkonia result~\cite{QWG}  from the Tevatron experiments 
is the ratio of direct production of $\chi_{c1}$ to $\chi_{c2}$ 
in $p \bar p$ collisions, or
\begin{equation}
\Rprod = \frac{\sigma(p\bar{p}\to\chi_{c1}X)}{\sigma(p\bar{p}\to\chi_{c2}X)}~.
\label{eqn:Rprod}
\end{equation}
This
is typically measured in the radiative decay of these $L=1$ states
to the $J/\psi$; {\it i.e.,} the experimentally accessible
quantity is $\Rprod\cdot R_{\gamma J/\psi}$, with
\begin{equation}
R_{\gamma J/\psi} = \frac
{{\cal B}(\chi_{c1}\to\gamma J/\psi)}
{{\cal B}(\chi_{c2}\to\gamma J/\psi)}~.
\label{eqn:ratio}
\end{equation}

The CDF Collaboration 
has a published result~\cite{CDFold} of
$\Rprod = 1.04 \pm 0.29 \pm 0.12$.
Models that expect $\Rprod$ to be the ratio of the available
spin states, such as ``color-evaporation'', predict
$\Rprod$ = 3/5; older, NRQCD predictions~\cite{theory} 
are for even smaller values.
The newest CDF preliminary measurement is $\Rprod\sim 1.4$~\cite{CDFQWG}, 
or roughly twice the spin-counting prediction. 
This CDF result promises to have statistical
uncertainties in $\Rprod$ of $\sim 6\%$ and systematic uncertainties
dominated by the lack of knowledge of $R_{\gamma J/\psi}$. 

Direct use of the individual Particle Data Group (PDG) 
values~\cite{PDG2006} for the two radiative branching
fractions in the numerator and denominator of
$R_{\gamma J/\psi}$ 
does not consider the possible cancelation of 
correlated experimental uncertainties. 
This note uses CLEO publications to obtain $R_{\gamma J/\psi}$,
taking such cancelations into account, thereby reducing
the total uncertainty on this ratio.

The 2006 PDG~\cite{PDG2006} values give
\begin{equation}
R_{\gamma J/\psi}^{PDG} = \frac
{0.356\pm 0.019}
{0.202\pm 0.010}
= 1.76 \pm 0.13~;
\label{eqn:RPDG}
\end{equation}
this has a 7.3\% relative uncertainty in $R_{\gamma J/\psi}$.  Note that
the 2006 PDG obtains the numerator and denominator from a global
fit which does include recent 
CLEO measurements~\cite{PRL,PRD}.\footnote{The corresponding 
value from the PDG in 2004~\cite{PDG2004}, which predated
Refs.~\cite{PRL,PRD}, was 
$R_{\gamma J/\psi}^{2004} = (31.6 \pm3.3)/(20.2 \pm 1.7) = 
1.56 \pm 0.32$, a 20\% assessment.}

CLEO has published the two photon cascade branching fractions
for $\psi {\rm(2S)}\to\gamma\chi_{cJ}\to\gamma\gamma J/\psi$ for the
three $J$ values in Ref.~\cite{PRL}.
These branching fractions
have noticable systematic uncertainties from $N_{2S}$, the number of
parent $\psi$(2S), and ${\cal B}_{\ell\ell}$ of the resulting
$J/ \psi$ state. 
These same two systematic uncertainties are present in the measurement,
again reported in Ref.~\cite{PRL}, of the di-pion
transition 
${\cal B}(\psi {\rm(2S)}\to \pi^{+}\pi^{-}J/\psi )$.  
The table in that publication
further gives the ratios of the two photon cascade branching fractions 
to this di-pion branching fraction;  such ratios have the
two above-mentioned systematics from $N_{2S}$ and ${\cal B}_{\ell\ell}$
canceling.  Copying directly from that table, we have:  
\begin{equation}
{\cal B}(\psi {\rm(2S)}\to\gamma\chi_{c1}\to\gamma\gamma J/\psi)/
 {\cal B}(\psi {\rm(2S)}\to \pi^{+}\pi^{-} J/\psi )
= 10.24 \pm 0.17 \pm 0.23
\label{eqn:ratio1}
\end{equation}
\begin{equation}
{\cal B}(\psi {\rm(2S)}\to\gamma\chi_{c2}\to\gamma\gamma J/\psi)/
 {\cal B}(\psi {\rm(2S)}\to \pi^{+}\pi^{-} J/\psi )
= 5.52 \pm 0.13 \pm 0.13
\label{eqn:ratio2}
\end{equation}

Ref.~\cite{PRL} asserts that there is a 0.75\%
systematic uncertainty for finding a photon pair and a 0.4\% 
uncertainty for finding each of the charged pions (0.8\% uncertainty
for the $\pi^{+}\pi^{-}$ pair).  Thus taking a ratio of Eqns.~\ref{eqn:ratio1}
and \ref{eqn:ratio2} allows cancelation of these contributions to the
uncertainty, once each in quadarture in the numerator and 
denominator.

To obtain a CLEO value for $R_{\gamma J/\psi}$ we also need the
ratio presented by CLEO in Ref.~\cite{PRD} for the transition
branching fractions from the $\psi$(2S) to the $\chi_{cJ}$ states,
namely
\begin{equation}
\frac{{\cal B}(\psi {\rm (2S)}\to\gamma\chi_{c2})}
{{\cal B}(\psi {\rm (2S)}\to\gamma\chi_{c1})} = 1.03 \pm 0.02 \pm 0.03~.
\label{eqn:TS}
\end{equation}

Our final result is then, using Eqns.~\ref{eqn:ratio1},
\ref{eqn:ratio2} and \ref{eqn:TS}:
\begin{equation}
\label{eqn:result}
R_{\gamma J/\psi}^{\mathrm{CLEO}} = 
\frac
{{\cal B}(\psi {\rm(2S)}\to\gamma\chi_{c1}\to\gamma\gamma J/\psi)/
 {\cal B}(\psi {\rm(2S)}\to \pi^{+}\pi^{-} J/\psi )}
{{\cal B}(\psi {\rm(2S)}\to\gamma\chi_{c2}\to\gamma\gamma J/\psi)/
 {\cal B}(\psi {\rm(2S)}\to \pi^{+}\pi^{-} J/\psi )} 
\cdot 
\frac
{{\cal B}(\psi {\rm (2S)}\to\gamma\chi_{c2})}
{{\cal B}(\psi {\rm (2S)}\to\gamma\chi_{c1})} 
$$
$$
= 1.91 \pm 0.10~.
\end{equation}
In Eqn.~\ref{eqn:result} we have combined
CLEO statistical and systematic uncertainties in quadrature; the
result represents a clear improvement in uncertainty over the value 
derived from the PDG listings.
  
We wish to thank Geoff Bodwin for his discussions on
the theoretical framework and importance of this result and
John Yelton for suggestions that improved the clarity of this note.  
This work in supported by the US National Science Foundation under
cooperative agreement PHY-0202078.
\vspace{1.0in}

\begin{thebibliography}{9}
%
\bibitem{QWG}
N. Brambilla \etal ~(Quarkonium Working Group),
{\it Heavy Quarkonium Physics},\\ 
hep-ph/0412158.
%
\bibitem{CDFold}
T.~Affolder \etal ~(CDF Collaboration), Phys. Rev. Lett. {\bf 86},
3963 (2001).
%
\bibitem{theory}
M. Beneke, hep-ph/9703429.
%
\bibitem{CDFQWG}
M.~J.~Kim, presentation and private communication
at the Int'l Workshop on Heavy Quarkonium, Brookhaven National Laboratory,
June 2006.
%
\bibitem{PDG2006}
W.-M. Yao \etal ~(Particle Data Group), J. Phys. G {\bf 33}, 1 (2006).
%
\bibitem{PRL}
N.~E.~Adam \etal ~(CLEO Collaboration), Phys Rev. Lett. {\bf 94},
232002 (2005).
%
\bibitem{PRD}
S.~B.~Athar \etal ~(CLEO Collaboration), Phys. Rev. D{\bf 70},
112002 (2004).
%
\bibitem{PDG2004}
S.~Eidelman \etal ~(Particle Data Group), Phys. Lett. {\bf B592}, 1 (2004).
%
\bibitem{Bmumu}
Z.~Li \etal ~(CLEO Collaboration),  Phys. Rev. D{\bf 71},
111103 (2005).
%
\end{thebibliography}
\end{document}